\documentclass[12pt]{iopart}

\usepackage{setstack}
\usepackage{iopams}
\usepackage{graphicx}% Include figure files
\usepackage{dcolumn}% Align table columns on decimal point
\usepackage{bm}% bold math
\usepackage[ansinew]{inputenc}
\usepackage{epsfig}
\usepackage{color}
\usepackage{units}

\usepackage{bbm}

\begin{document}

\newcommand{\spinup}{|\uparrow\rangle}
\newcommand{\spindown}{|\downarrow\rangle}
\newcommand{\bess}[1]{{J_{#1}}}
\renewcommand{\vec}[1]{\ensuremath{\boldsymbol{#1}}}
\newcommand{\Ca}{$^{40}$Ca$^+$}
\newcommand{\bra}[1]{\ensuremath{\langle #1|}}   %defines a bra
\newcommand{\ket}[1]{\ensuremath{|#1\rangle}}   %defines a ket
\newcommand{\qd}{\ket{\mbox{$\downarrow$}}}
\newcommand{\qu}{\ket{\mbox{$\uparrow$}}}
\newcommand{\qdd}{\ket{\mbox{$\downarrow \downarrow$}}}
\newcommand{\qud}{\ket{\mbox{$\uparrow \downarrow$}}}
\newcommand{\qdu}{\ket{\mbox{$\downarrow \uparrow$}}}
\newcommand{\quu}{\ket{\mbox{$\uparrow \uparrow$}}}
\newcommand{\todo}[1]{\marginparwidth=1.0cm \marginparsep=0.1cm\marginpar{%
    \null\vspace*{-1.1\baselineskip}%
    \rule[0pt]{1.cm}{1pt}\\
    \sffamily\tiny#1\\
    \rule[2pt]{1.cm}{1pt}}}

\article[]{}{Deterministic entanglement of ions in thermal states
of motion}

\author{G~Kirchmair$^{1,2}$, J~Benhelm$^{1,2}$, F~Z{\"a}hringer$^{1,2}$, R~Gerritsma$^{1,2}$, C~F~Roos$^{1,2}$ and R~Blatt$^{1,2}$}

\address{$^1$Institut f\"ur
Quantenoptik und Quanteninformation, \"Osterreichische Akademie
der Wissenschaften, Otto-Hittmair-Platz 1, A-6020 Innsbruck,
Austria}
\address{$^2$Institut f\"ur Experimentalphysik, Universit\"at Innsbruck,
Technikerstr.~25, A-6020 Innsbruck, Austria}

\ead{Christian.Roos@uibk.ac.at}

\date{\today}% It is always \today

\begin{abstract}
We give a detailed description of the implementation of a
M{\o}lmer-S{\o}rensen gate entangling two \Ca\ ions using a
bichromatic laser beam near-resonant with a quadrupole transition.
By amplitude pulse shaping and compensation of AC-Stark shifts we
achieve a fast gate operation without compromising the error rate.
Subjecting different input states to concatenations of up to $21$
individual gate operations reveals Bell state fidelities above
0.80. In principle, the entangling gate does not require ground
state cooling of the ions as long as the Lamb-Dicke criterion is
fulfilled. We present the first experimental evidence for this
claim and create Bell states with a fidelity of 0.974(1) for ions
in a thermal state of motion with a mean phonon number of
$\bar{n}=20(2)$ in the mode coupling to the ions' internal states.
\end{abstract}

%\pacs{??}
\submitto{\NJP}
%\maketitle

%\tableofcontents

\section{Introduction}
Building a device that is able to carry out arbitrary calculations
by exploiting the laws of quantum physics has been an experimental
challenge for more than a decade now. A large variety of physical
implementations have been conceived to meet the requirements for
quantum information processing summarized
in~\cite{DiVincenzo2000a}. Among these implementations, strings of
ions stored in linear Paul traps and manipulated by laser pulses
have proven to be a particularly successful architecture to
realize quantum information processing. Experiments with trapped
ions have shown long relevant coherence times
\cite{Langer2005,olmschenk2007,Benhelm:2008c}, the ability to
faithfully initialize and read-out qubits \cite{Roos2006,
Myerson:2008a} and high fidelity quantum operations
\cite{Knill:2007,Leibfried2003a,Benhelm:2008b}. Current efforts
are focussed on scaling up ion trap experiments to handle many
ions, improving the quality and speed of the basic operations and
integrating the various techniques into a single system.
Concerning the basic operations, the realization of universal
multi-qubit gates is particularly challenging. Many different
types of gates have been proposed over the last years and several
of them have been experimentally investigated. Gates using a
collective interaction \cite{Leibfried2003a,Sackett2000} between
the ions and the laser field - until recently only applied to
qubits encoded in the hyperfine structure (hyperfine qubits) -
were very successful in creating multi-particle entangled states
and demonstrating simple quantum error correction techniques
\cite{Chiaverini:2004a}.

Recently we demonstrated the first application of a
M{\o}lmer-S{\o}rensen gate operation to an optical qubit, i.e. a
qubit encoded in a ground and a metastable state of \Ca\ ions,
deterministically creating Bell states with a so far unmatched
fidelity of 0.993(1) \cite{Benhelm:2008b}. Here, we present a
further investigation of this universal gate operation acting on
optical qubits and extend the theoretical and experimental
analysis. Particular emphasis is put on the compensation of
AC-Stark shifts and amplitude pulse shaping to reach high
fidelities without compromising the gate speed substantially. The
gate characterization is extended further by investigating the
fidelity decay for different input states after up to $21$
individual operations.

Moreover, we report on the first experiments demonstrating a
universal entangling gate operating on Doppler-cooled ions. We
derive simple expressions~\cite{Roos:2008a} for the qubit
populations under the action of the gate and and use these
equations to infer the mean vibrational quantum number $\bar{n}$
of the axial center-of-mass mode. For ions in a thermal state with
$\bar{n}=20(2)$, we obtain Bell states with a fidelity of
$0.974(1)$.

The ability to implement high fidelity multi-qubit operations on
Doppler-cooled ions is of practical interest in ion trap quantum
information processing as the implementation of quantum algorithms
demands several techniques that do not conserve the ions'
vibrational quantum state: (i) State detection of ancilla qubits
as required by quantum error correction schemes~\cite{Steane:1998}
can excite the ion string to a thermal motional state close to the
Doppler limit because of the interaction with the laser inducing
the ions to fluoresce. (ii) Experiments with segmented traps
structures where ion strings are split into smaller strings also
tend to heat up the ions slightly ~\cite{Rowe2002}. Here,
the availability of high-fidelity gate operations even
for thermal states may provide a viable alternative
to the technically involved re-cooling techniques
using a different ion species~\cite{Blinov2002,Barrett2003}.

\section{\label{gate mechanism}Theoretical gate description}

\subsection{M{\o}lmer-S{\o}rensen gate}
\begin{figure}
\centering
\includegraphics[width=140mm]{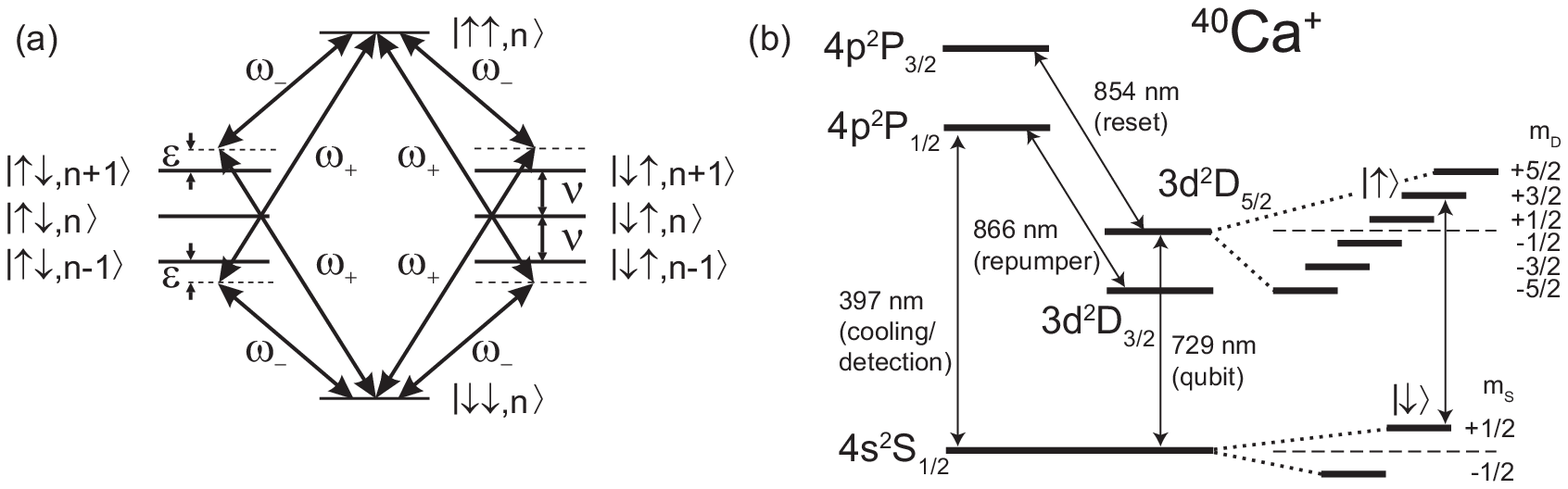}
\caption{(a) M{\o}lmer-S{\o}rensen interaction scheme. A
bichromatic laser field couples the qubit states $\qdd
\leftrightarrow\quu$ via the four interfering paths shown in the
figure. Similar processes couple the states $\qud
\leftrightarrow\qdu$. The frequencies $\omega_\pm$ of the laser
field are tuned close to the red and the blue motional sidebands
of the qubit transition with frequency $\omega_0$, and satisfy the
resonance condition $2\omega_0=\omega_++\omega_-$. The vibrational
quantum number is denoted $n$. (b) Level scheme of \Ca showing the
transitions used for cooling/detecting, repumping and resetting
the state of the ion as well as the qubit transition. The qubit is
encoded in the metastable state $\qu=|D_{5/2}, m=3/2\rangle$ and
the ground state $\qd=|S_{1/2},m=1/2\rangle$.
%Additionally the finestructure splitting of the $D_{5/2}$ and the
%$S_{1/2}$ states is shown.
} \label{fig:MSgateMechanism}
\end{figure}

A two-qubit quantum gate that is equivalent to a controlled-NOT
gate up to local operations is achieved by the action of a Hamiltonian
$H\propto \sigma_n\otimes\sigma_n$, where
$\sigma_n=\mathbf{\sigma\cdot n}$ is a projection of the vector of
Pauli spin matrices onto the direction $\mathbf n$~\cite{Lee2005}.
Two prominent examples of this type of gate are the conditional
phase gate~\cite{Leibfried2003a,Milburn2000} and the
M{\o}lmer-S{\o}rensen gate~\cite{Soerensen1999,Solano1999,Sackett2000}. In the latter case,
correlated spin flips between the states
$\qu\qu\leftrightarrow\qd\qd $ and $\qu\qd\leftrightarrow\qd\qu$
are induced by a Hamiltonian
\begin{equation}
\label{sigmaphisigmaphi} H\propto
\sigma_\phi\!\otimes\!\sigma_\phi\;\;\;\;\mbox{where}\;\;\;\;
\sigma_\phi=\cos\phi\sigma_x+\sin\phi\sigma_y.
\end{equation}
The unitary operation
$U=\exp(i\frac{\pi}{4}\sigma_\phi\otimes\sigma_\phi)$ maps product
states onto maximally entangled states. In 1999, the proposal was
made to realize an effective
Hamiltonian~\cite{Soerensen1999,Solano1999} taking the form
(\ref{sigmaphisigmaphi}) by exciting both ions simultaneously with
a bichromatic laser beam with frequencies
$\omega_\pm=\omega_0\pm\delta$ where $\omega_0$ is the qubit
transition frequency and $\delta$ is close to a vibrational mode
of the two-ion crystal with frequency $\nu$ (see
\fref{fig:MSgateMechanism}(a)). Changing into an interaction picture and
performing a rotating-wave approximation, the time-dependent
Hamiltonian
\begin{equation}
\label{FullHamiltonian} H(t)=\hbar\Omega(e^{-i\delta t}+e^{i\delta
t})e^{i\eta(ae^{-i\nu t}+a^\dagger e^{i\nu
t})}(\sigma_+^{(1)}+\sigma_+^{(2)})+\mbox{h.c.}
\end{equation}
 is well approximated by
\begin{equation}
\label{HMS}
 H(t)=-\hbar\eta\Omega(a^\dagger e^{i(\nu-\delta)t}+a
e^{-i(\nu-\delta)t})S_y
\end{equation}
in the Lamb-Dicke regime where the Lamb-Dicke factor $\eta$
satisfies the condition $\eta\ll 1$. In~(\ref{HMS}), we use a
collective spin operator $S_y=\sigma_y^{(1)}+\sigma_y^{(2)}$ and
denote the laser detuning from the motional sidebands by
$\nu-\delta=\epsilon$. The Rabi frequency on the carrier
transition is denoted $\Omega$, and $a,a^\dagger$ are the phonon
annihilation and creation operators, respectively. This
Hamiltonian can be exactly integrated~\cite{Soerensen2000}
yielding the propagator
\begin{equation}
U(t)=\hat{D}(\alpha(t)S_y)\exp\left(i(\lambda t -\chi\sin(\epsilon
t))S_y^2\right),\label{MSPropagator}
\end{equation}
where $\alpha(t)=\frac{\eta\Omega}{\epsilon}(e^{i\epsilon t}-1)$,
$\lambda=\eta^2\Omega^2/\epsilon$,
$\chi=\eta^2\Omega^2/\epsilon^2$, and $\hat{D}(\alpha)=e^{\alpha
a^\dagger-\alpha^\ast a}$ is a displacement operator. For a gate
time $t_{\rm gate}=2\pi/|\epsilon|$, the displacement operator
vanishes so that the propagator $U(t_{\rm gate})=\exp(i\lambda
t_{\rm gate}S_y^2)$ can be regarded as being the action of an
effective Hamiltonian
\begin{equation}
\label{Heff} H_{\rm eff}=-\hbar\lambda
S_y^2=-2\hbar\lambda(\mathbbm{1}+\sigma_y\otimes\sigma_y)
\end{equation}
taking on the form given in (\ref{sigmaphisigmaphi}). Setting
$\Omega=|\epsilon|/(4\eta)$, a gate is realized capable of
maximally entangling ions irrespective of their motional state.
\begin{figure}
\centering
\includegraphics[width=100mm]{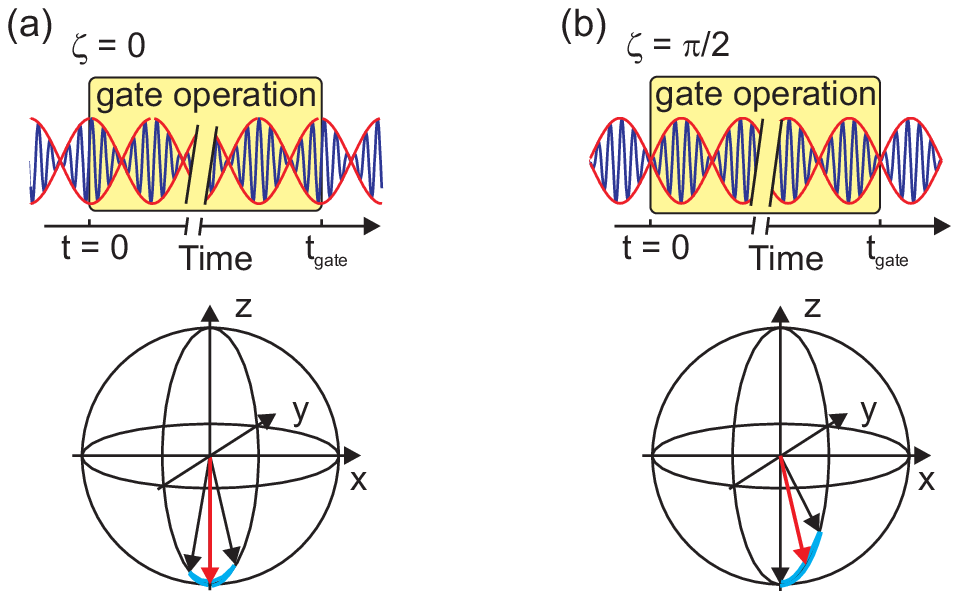}
\caption{\label{fig:blochspherepicture} Effect of non-resonant
excitation of the carrier transition. (a) For $\zeta=0$,
the gate starts at a maximum of the intensity-modulated beam. In
this case, a Bloch vector initially centered at the south pole of
the Bloch sphere performs oscillations that are symmetric around
the initial position. (b) For $\zeta=\pi/2$, the gate starts at
the minimum of the intensity modulation. In this case, the average
orientation of the Bloch vector is tilted with respect to its
initial position.}
\end{figure}
In the description of the gate mechanism given so far, a coupling
of the light field to the carrier transition was neglected based
on the assumption that the Rabi frequency $\Omega$ was small
compared to the detuning $\delta$ of the laser frequency
components from the transition. In this case, small non-resonant
Rabi oscillations that appear on top of the gate dynamics are the
main effect of coupling to the carrier transition. Since a
maximally entangling gate requires a Rabi frequency
$\Omega\propto\eta^{-1}t_{\rm gate}^{-1}$, the question of whether
$\Omega\ll\delta$ holds becomes crucial in the limit of fast gate
operations and small Lamb-Dicke factors. Our
experiments~\cite{Benhelm:2008b} are exactly operating in this
regime, and it turns out that non-resonant excitation of the
carrier transition has further effects beyond inducing
non-resonant oscillations~\cite{Roos:2008a}. This becomes apparent
by interpreting terms in the Hamiltonian in a different way: The
red- and blue-detuned frequency components
$E_\pm=E_0\cos((\omega_0\pm\delta)t\pm\zeta)$ of equal intensity
may be viewed as a single laser beam
$E(t)=E_++E_-=2E_0\cos(\omega_0 t)\cos(\delta t+\zeta)$ that is
resonant with the qubit transition but amplitude-modulated with
frequency $\delta$. Here, the phase $\phi$ which determines
whether the gate operation starts in a maximum ($\zeta=0$) or a
minimum ($\zeta=\pi/2$) of the intensity of the
amplitude-modulated beam has a crucial influence on the gate. This
can be intuitively understood by considering the initial action
the gate exerts on an input state in the Bloch sphere picture
shown in \fref{fig:blochspherepicture}. For short times, coupling
to the sidebands can be neglected which justifies the use of a
single-ion picture. The dynamics is essentially the one of two
uncoupled qubits. The fast dynamics of the gate is induced by
excitation of the ions on the carrier transition. For $\zeta=0$,
the Bloch vector of an ion initially in state $\qd$ will oscillate
with frequency $\delta$ along a line centered on the south pole of
the Bloch sphere. For $\zeta=\pi/2$, the oscillation frequency is
the same, however, the time-averaged position of the Bloch vector
is tilted by an angle
\begin{equation}
\psi=\frac{4\Omega}{\delta}\sin\zeta \label{psi}
\end{equation}
with respect to the initial state $\qd$. This effect has a
profound influence on the gate action. A careful analysis of the
gate mechanism~\cite{Roos:2008a} taking into account the
non-resonant oscillations reveals that the Hamiltonian (\ref{HMS})
is changed into
\begin{equation}
\label{HMSmod}
 H(t)=-\hbar\eta\Omega(a^\dagger e^{i(\nu-\delta)t}+a
e^{-i(\nu-\delta)t})S_{y\!\,,\psi}\,,
\end{equation}
where
\begin{equation}
S_{y\!\,,\psi} = S_y\cos\psi  + S_z\sin\psi\,,
\end{equation}
and that the propagator (\ref{MSPropagator}) needs to be replaced
by
\begin{equation}
U(t)=\exp(-iF(t)S_x)\hat{D}(\alpha(t)S_{y\!\,,\psi})\exp\left(i(\lambda
t -\chi\sin(\epsilon
t))S_{y\!\,,\psi}^2\right),\label{MSPropagatorMod}
\end{equation}
where the term containing $F(t)=\frac{2\Omega}{\delta}(\sin(\delta
t+\zeta)-\sin\zeta)$ describes non-resonant excitation of the
carrier transition. The dependence of the propagator on the exact
value of $\zeta$ is inconvenient from an experimental point of
view. To realize the desired gate, precise control over $\zeta$ is
required. In addition, the gate duration must be controlled to
better than a fraction of the mode oscillation period because of
the non-resonant oscillation. Fortunately, both of these problems
can be overcome by shaping the overall intensity of the laser
pulse such that the Rabi frequency $\Omega(t)$ smoothly rises
within a few cycles $2\pi/\delta$ to its maximum value
$\Omega_{\rm gate}\approx|\epsilon|/(4\eta)$ and smoothly falls
off to zero at the end of the gate. In this case, the non-resonant
oscillations vanish and~(\ref{psi}) shows that the operator
$S_{y\!\,,\psi}(t)$ adiabatically follows the laser intensity so
that it starts and ends as the desired operator $S_y$ irrespective
of the phase $\zeta$. For intensity-shaped pulses, the propagator
(\ref{MSPropagator}) provides therefore an adequate description of
the gate action.

\subsection{AC-Stark shifts\label{AC_stark}}

In the description of the gate mechanism given so far the ion was treated
as an ideal two-level system. AC-Stark shifts are completely
insignificant provided that the intensities of the blue- and the
red-detuned frequency components are the same since in this case
light shifts of the carrier transition caused by the blue-detuned
part are exactly canceled by light shifts of the red-detuned light
field. Similarly, light shifts of the blue-detuned frequency
component non-resonantly exciting the upper motional sideband are
perfectly canceled by light shifts of the red-detuned frequency
component coupling to the lower motional sideband.

For an experimental implementation with calcium ions, we need to
consider numerous energy levels (see
\fref{fig:MSgateMechanism}(b)). Here, the laser field inducing the
gate action causes AC-Stark shifts on the qubit transition
frequency due to non-resonant excitation of far-detuned dipole
transitions and also of other $S_{1/2}\leftrightarrow D_{5/2}$
Zeeman transitions. The main contributions arise from couplings
between the qubit states and the $4p-$states that are mediated by
the dipole transitions $S_{1/2}\leftrightarrow P_{1/2}$,
$S_{1/2}\leftrightarrow P_{3/2}$, $D_{5/2}\leftrightarrow
P_{3/2}$. Other transitions hardly matter as can be checked by
comparing the experimental results obtained
in~\cite{Haeffner2003a} with numerical results based on the
transition strengths~\cite{James1998} of the dipole transitions
coupling to the $4p-$states. For suitably choosen k-vector and
polarization of the bichromatic laser beam, these shifts are
considerably smaller than the strength $\lambda$ of the gate
interaction.

AC-Stark shifts can be compensated for by a suitable detuning of
the gate laser. An alternative strategy consists in introducing an
additional AC-Stark shift of opposite sign that is also caused by
the gate laser beam~\cite{Haeffner2003a}. This approach has the
advantage of making the AC-Stark compensation independent of the
gate laser intensity. In contrast to previous gates relying on
this technique~\cite{Schmidt-Kaler2003a} where the AC-Stark shift was
caused by the quadrupole transition and compensated by coupling to
dipole transitions, here, the AC-Stark shift is due to dipole transitions
and needs to be compensated by coupling to the quadrupole transition.

For ions prepared in the ground state of motion ($n=0$), a
convenient way of accomplishing this task is to perform the gate
operation with slightly imbalanced intensities of the blue- and
the red-detuned laser frequency components. Setting the Rabi
frequency of the blue-detuned component to $\Omega_{\rm
b}=\Omega(1+\xi)$ and the one of the red-detuned to $\Omega_{\rm
r}=\Omega(1-\xi)$, an additional light shift caused by coupling to
the carrier transition is induced that amounts to
$\delta^{(C)}_{\rm ac}=2(\Omega_{\rm r}^2-\Omega_{\rm
b}^2)/\delta=-8\Omega^2\xi/\delta$. Now, the beam imbalance
parameter $\xi$ needs to be set such that the additional light
shift exactly cancels the phase shift $\phi=\delta_{\rm ac}t_{\rm
gate}$ induced by the dipole transitions during the action of the
gate. Taking into account that $t_{\rm gate}=2\pi/\epsilon$ and
$\Omega=|\epsilon|/(4\eta)$, this requires
$\xi=(\delta\eta^2/|\epsilon|)(\phi/\pi)$.

Apart from introducing light shifts, setting $\xi\neq 0$ also
slightly changes the gate Hamiltonian (\ref{Heff}) from $H_{\rm
eff}=-\lambda S_y^2$ to $H_{\rm eff}=-\lambda(S_y^2+\xi^2S_x^2)$
\cite{Unanyan:2003} since the coupling between the states \qdd,
\quu\ is proportional to $2\Omega_{\rm b}\Omega_{\rm
r}=2\Omega^2(1-\xi^2)$ whereas the coupling between \qdu, \qud\ is
proportional to $\Omega_{\rm b}^2+\Omega_{\rm
r}^2=2\Omega^2(1+\xi^2)$. However, as long as $\xi\ll 1$ holds --
which is the case in the experiments described in the next section
-- this effect is extremely small as the additional term is only
quadratic in $\xi$.

Another side effect of setting $\xi\neq 0$ is the occurrence of an
additional term $\propto S_z a^\dagger a$ in the Hamiltonian. It
is caused by AC-Stark shifts arising from coupling to the upper
and lower motional sideband which no longer completely cancel each
other. The resulting shift of the qubit transition frequency
depends on the vibrational quantum number $n$ and is given by
$\delta^{\rm (SB)}=(8\eta^2\Omega^2/\epsilon)\xi n=(\epsilon/2)\xi
n$. Simulations of the gate action based on
(\ref{FullHamiltonian}) including an additional term $\propto S_z$
accounting for AC-Stark shifts of the dipole transitions and
power-imbalanced beams show that the unwanted term $\propto S_z
a^\dagger a$ has no severe effects for ions prepared in the
motional ground state as long as $\xi\ll 1$. However, for ions in
Fock states with $n>0$, this is not the case. Taking the parameter
set $\xi=0.05$, $\nu=(2\pi)\,1230$~kHz, and
$\epsilon=(2\pi)\,20$~kHz as an example, the following results are
obtained: applying the gate to ions prepared in
$|\downarrow\downarrow\rangle|n=0\rangle$, a Bell state is created
with fidelity $0.9996$. For $n=1$, the fidelity drops to $0.992$,
and for $n=2$ to even $0.976$. This loss of fidelity can be only
partially recovered by shifting the laser frequency by
$\delta^{\rm (SB)}$, the resulting fidelity being $0.996$ and
$0.988$, respectively. For higher motional states, the effect is
even more severe and shows that this kind of AC-Stark compensation
is inappropriate when dealing with ions in a thermal state of
motion with $\bar{n}\gg 1$. Instead of compensating the AC-Stark
shift by imbalancing the beam powers, in this case, the laser
frequency needs to be adjusted accordingly.

\section{Experimental setup}

\begin{figure}
\centering
\includegraphics[width=120mm]{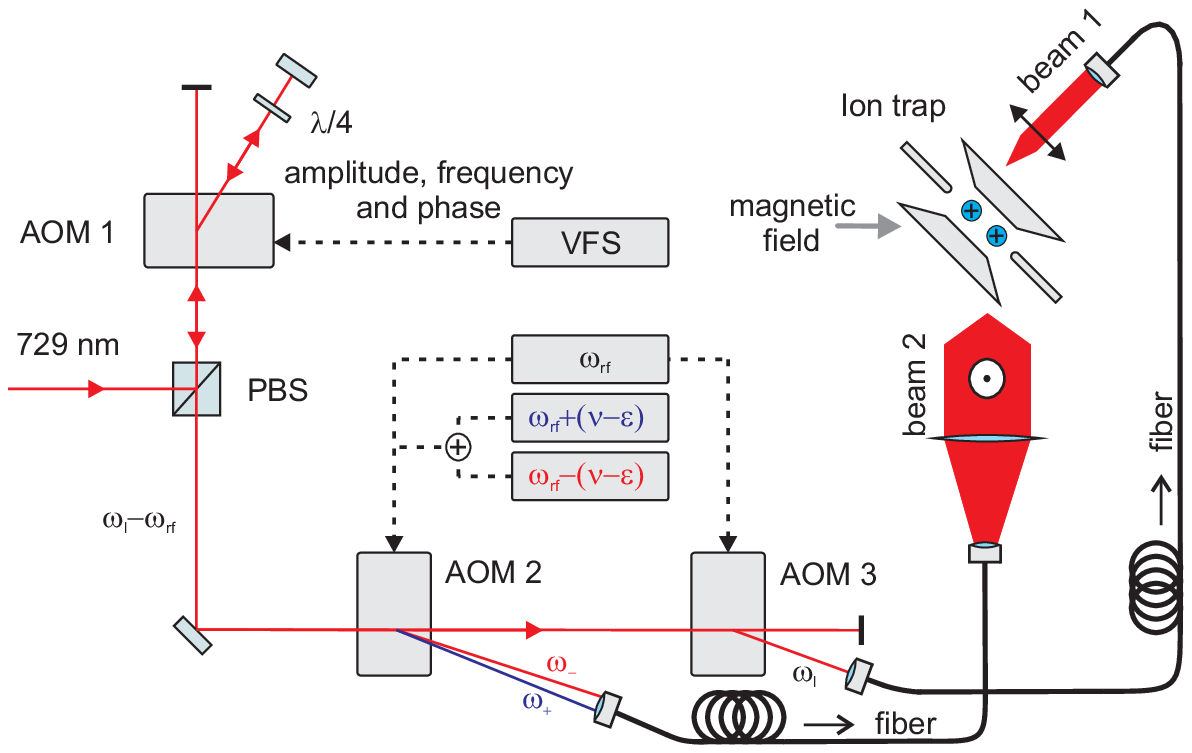}
\caption{\label{fig:bichrom_beam} Trap geometry and optical qubit
light field generation. AOM~1 controls the overall frequency
$\omega_l$ and the amplitude of the laser beams at \unit[729]{nm}.
AOM~2 is used to switch beam~2 which illuminates both ions
simultaneously. When supplied with two frequencies $\omega_{\rm
rf}\pm \delta$ where $\omega_{\rm rf}/(2\pi)=\unit[80]{MHz}$, it
creates a bichromatic light field in the first order of
diffraction. AOM~3 switches beam~1 which addresses only one of the
ions. Both beams are guided with single mode optical fibers to the
trap. The geometric alignment of polarizations, trap axis and
magnetic field is as sketched. Radio-frequency signals supplying
the AOMs are indicated as dotted lines.}
\end{figure}

Two \Ca\ ions are stored in a linear Paul trap with an axial trap
frequency $\nu/(2\pi) =  \unit[1.232]{MHz}$ corresponding
to an inter-ion distance of \unit[5]{$\mu$m}. The \Ca\ optical qubit
consists of the metastable state $\qu=|D_{5/2}, m=3/2\rangle$ with a
lifetime of \unit[1.17]{s} and the ground state $\qd=|S_{1/2},
m=1/2\rangle$ (see \fref{fig:MSgateMechanism}(b)).
These two energy levels are connected via a quadrupole
transition at a wavelength of \unit[729]{nm}. Laser light at
\unit[397]{nm} is used for Doppler-cooling and state detection on
the $S_{1/2}\leftrightarrow P_{1/2}$ transition with an additional
repumping laser at \unit[866]{nm} on the $D_{3/2}\leftrightarrow
P_{1/2}$ transition. Fluorescence light is detected by means of a
photomultiplier tube. For two ions we discriminate between
$\quu$, $\{\qud\, \mbox{or}\,\qdu\}$ and $\qdd$, the populations of
which are labeled by $p_0$, $p_1$, and $p_2$ according to the
number of ions fluorescing.

A titanium sapphire laser~\cite{Benhelm2007}, whose frequency is
stabilized to a high finesse Fabry-Perot cavity
\cite{Notcutt2005}, is used to excite the quadrupole transition
for sideband-cooling, frequency-resolved optical pumping and
performing quantum logic operations. Frequency drifts of maximally
\unit[3]{Hz/s} induced by the reference cavity are canceled by an
automated measurement routine referencing the laser frequency to
the optical qubit transition frequency and detecting the
magnitude of the magnetic field at the ions' location of about \unit[4]{Gauss}.

The setup for controlling the laser driving the qubit transition
is depicted in \fref{fig:bichrom_beam}. Laser light of
\unit[729]{nm} is sent to the ions from either of two directions,
each beam having a maximum light power of \unit[50]{mW}. Only
when single-ion addressing is required we use laser beam~1 focused
to a beam waist of \unit[3]{$\mu $m} at the trap center with a
$k$-vector perpendicular to the trap axis and a polarization that
couples to all possible transitions. All other operations are
accomplished with laser beam~2 whose $k$-vector encloses a
\unit[45]{°} angle with the axis of the ion string and is perpendicular to the
quantization axis defined by the direction of the magnetic field.
With a beam waist of \unit[14]{$\mu$m} at the trap center, this
beam is adjusted to illuminate both ions with equal intensity. The
polarization of this beam is set such that the coupling is maximal
for $\Delta m = \pm 1$ transitions whereas it vanishes for all
other transitions. The amplitude, frequency and phase of both
beams is controlled by the acousto-optical modulator~(AOM)~1 which
is driven by a versatile frequency source (VFS).
%(see\fref{fig:bichrom_beam})
Amplitude pulse shaping is achieved with a variable gain amplifier
controlled by a field programmable gate array. All radio-frequency
sources are phase-locked to an ultra-stable quartz oscillator. By
triggering each experimental cycle to the AC-power line we largely
reduce distortions caused by the \unit[1]{mG} magnetic field
fluctuations at \unit[50]{Hz}.

The AOMs~2 and 3 are used as switches for laser beams 1 and 2
applied from different directions. The bichromatic light field with
frequencies $\omega_{\pm}=\omega_0\pm\delta/(2\pi)$ is created by
driving AOM~2 simultaneously with the two frequencies $\omega_{\rm
rf}\pm\delta$, where $\delta=\nu-\epsilon$. A frequency
difference of $2\delta/(2\pi)=\unit[2.4]{MHz}$ leads to a diffraction into slightly
different directions with an angular separation as small as
\unit[0.025]{°} such that the coupling efficiency to the single
mode fiber is reduced by about $15\%$ compared to a single
frequency beam where AOM~2 is driven with $\omega_{\rm rf}$. To generate the
collective $\pi/2$-pulses needed for analyzing the gate action,
AOM~2 is driven with a single frequency $\omega_{\rm rf}$. A more
detailed description of the apparatus is given in
\cite{Benhelm:2008c}.

\section{Measurement results}

\begin{figure}
\centering
\includegraphics[width=140mm]{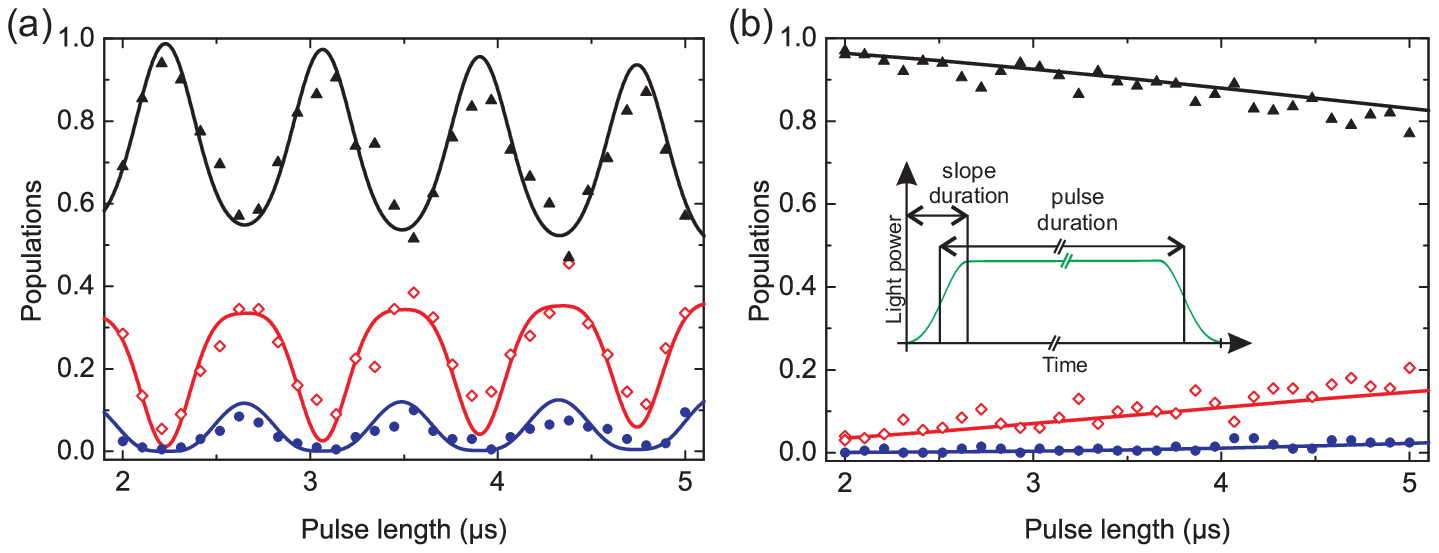}
\caption{\label{fig:offresexc} Effect of amplitude pulse shaping
on non-resonant population transfer caused by a bichromatic light
field non-resonantly exciting the carrier transition. Experimental
results are presented for a gate duration of $t_{\rm
gate}=\unit[25]{\mu s}$. A comparison of the evolution of the
populations $p_0 (\blacktriangle), p_1
(\textcolor[rgb]{1.00,0.00,0.00}{\diamond}),
p_2(\textcolor[rgb]{0.00,0.00,1.00}{\bullet}) $ for a
square-shaped pulse (a) with an amplitude-shaped pulse (b) shows a
suppression of the strong non-resonant oscillations for the latter
case. The slopes are shaped as a Blackman window with a duration
of \unit[2.5]{$\mu s$}, the figure inset showing the definitions
of pulse and slope duration. Numerical simulations suggest that
the actual pulse shape is not so important as long as the
switching occurs sufficiently slowly. The solid lines are
calculated from~(\ref{MSPropagatorMod}) and~(\ref{MSPropagator}).}
\end{figure}

The coupling strength of the laser to the qubit is calibrated by
recording resonant Rabi oscillations on the qubit transition. In
case of short gate operations, the large intensities lead to big
AC-Stark shifts and saturation of the gate coupling strength
\cite{Roos:2008a} which in turn necessitate a fine-adjustment of
the laser frequency and power.

\subsection{\label{shaping}Amplitude pulse shaping}

The merits of amplitude pulse shaping were studied by observing
the time evolution of the populations $p_i$ at the beginning of
the gate operation when the population transfer is dominated by
fast non-resonant coupling to the carrier. \Fref{fig:offresexc}
(a) shows the population evolution for the first $\unit[5]{\mu s}$
of a $\unit[25]{\mu s}$ gate operation based on a rectangular
pulse shape. Averaging over a randomly varying phase $\zeta$, we
observe strong oscillations with a period of $2\pi/\delta=\unit[0.84]{\mu s}$.
Panel (b) shows that the non-resonant excitations vanish
completely after application of amplitude pulse shaping with a
slope duration of $\unit[2.5]{\mu s}$ corresponding to three
vibrational periods of the center-of-mass mode. The slopes were shaped as a Blackman
window \cite{Harris:1978}, where the form of the shape is chosen such
that a shaped and a rectangular pulse of the same duration have
the same pulse area (see inset of panel (b)). Different pulse
lengths are achieved by varying the duration of the central time
interval during which the laser power is constant. The solid lines in the figure are calculated
from~(\ref{MSPropagatorMod}) and~(\ref{MSPropagator}) and fit the data well.

\subsection{AC-Stark shift compensation}

\begin{figure}
\centering
\includegraphics[width=140mm]{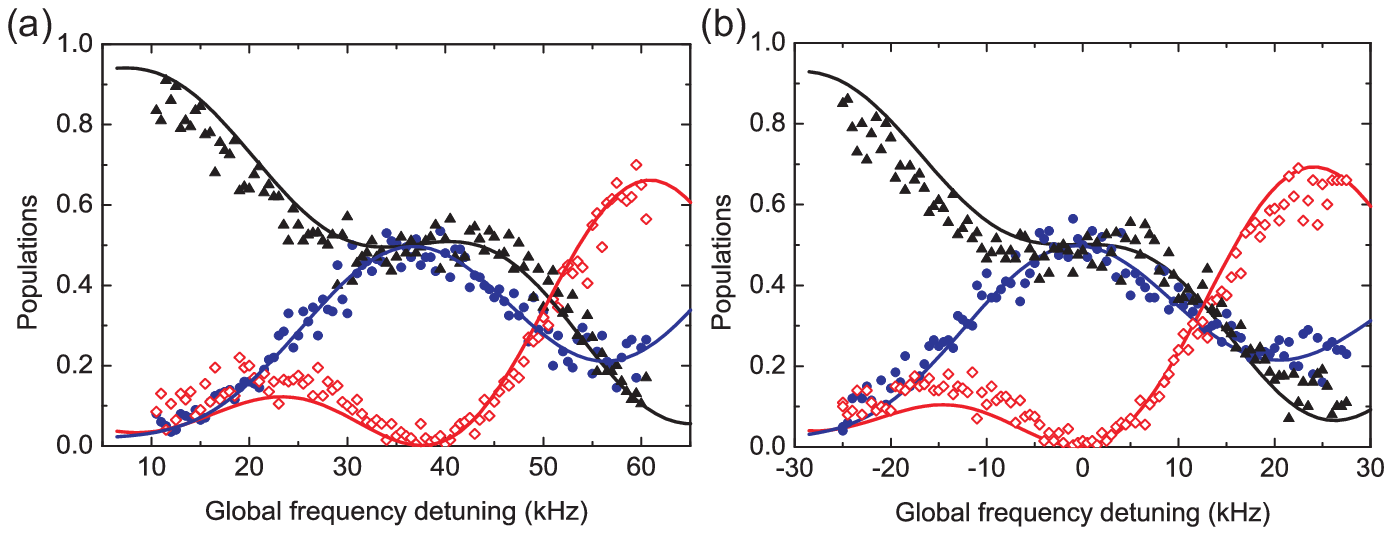}
\caption{\label{fig:frequ_scan} (a) Populations $p_0
(\blacktriangle), p_1 (\textcolor[rgb]{1.00,0.00,0.00}{\diamond}),
p_2(\textcolor[rgb]{0.00,0.00,1.00}{\bullet}) $ after a single
gate operation ($t_{\rm gate}=\unit[25]{\mu s}$) where the global
frequency detuning of the bichromatic entangling pulse is varied
by scanning AOM~1. A maximally entangled state is achieved for a
global frequency detuning of $\unit[(2\pi)\,37]{kHz}$ relative to
the qubit transition frequency due to AC-Stark shifts. (b)
Introduction of a beam imbalance $\xi = 0.08$ shifts the pattern
of the populations by the required amount to fully compensate for
the AC-Stark-shift (note the different x-axis offsets in (a) and
(b)). The solid lines are calculated by solving the
Schr{\"o}dinger equation for the Hamiltonian given in
(\ref{FullHamiltonian}) amended by a term accounting for the
measured AC-Stark shift.}
\end{figure}

The AC-Stark shift caused by bichromatic light with spectral
components each having a Rabi frequency of
$\Omega/(2\pi)=\unit[220]{kHz}$ (for $t_{\rm gate}=\unit[25]{\mu
s}$) is measured by scanning the global laser frequency using
AOM~1. The resulting populations after a gate operation are
depicted in \fref{fig:frequ_scan}~(a). We observe a drop of the
population $p_1$ to zero at a detuning of $\unit[(2\pi)\,37]{kHz}$
from the carrier transition. At this setting the ions are
maximally entangled. By changing the relative power of the
bichromatic field's frequency components such that $\xi = 0.08$
the AC-Stark shift is compensated. This translates the whole
excitation pattern in frequency space as can be seen in
\fref{fig:frequ_scan}~(b).

\begin{figure}
\centering
\includegraphics[width=80mm]{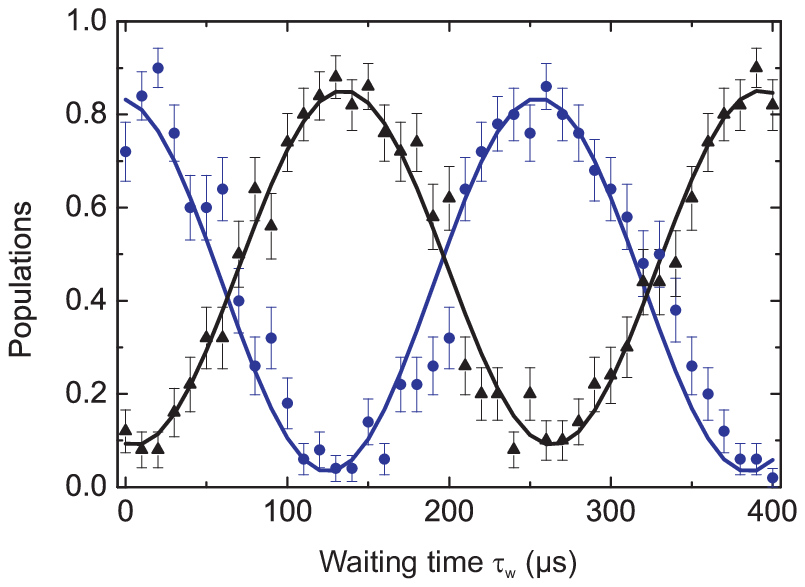}
\caption{\label{fig:MSramsey} Population evolution of $\quu
(\textcolor[rgb]{0.00,0.00,1.00}{\bullet})$ and $\qdd
(\blacktriangle)$ when scanning the waiting time between two
$\unit[25]{\mu s}$ gate pulses in a Ramsey-like experiment. For
this scan the detuning $\epsilon$ was set to $\unit[(2\pi)
40]{kHz}$. In this set of data, the AC-Stark shift was only
partially compensated by imbalancing the power of the two
frequency components. From the sinusodial fits shown as solid
lines, we infer an oscillation period of $\unit[258(4)]{\mu s}$
corresponding to a residual AC-Stark shift of $\unit[(2\pi)
1.94(3)]{kHz}$.}
\end{figure}

A more sensitive method to infer the remaining AC-Stark shift
$\delta_{AC}$ after a coarse pre-compensation consists in
concatenating two gates separated by a waiting time $\tau_{w}$ in
a pulse sequence akin to a Ramsey-type
experiment~\cite{Leibfried2004} and scanning $\tau_{w}$. This
procedure maps $\delta_{AC}$ to a phase $\phi=\delta_{AC}\tau_w$
which is converted into a population change $p_2=\cos^2(\phi)$,
$p_0=\sin^2(\phi)$ by the second gate pulse. For two ions, the
corresponding Ramsey pattern displayed in \fref{fig:MSramsey}
shows oscillations of the populations $p_0$ and $p_2$ with a
frequency of two times the remaining AC-Stark shift.

\subsection{Gate analysis}

A full characterization of the gate operation could be achieved by
quantum process tomography~\cite{Riebe:2006}. At present, however,
the errors introduced by single ion addressing and individual
qubit detection are on the few percent level in our experimental
setup which renders the detection of small errors difficult in the
entangling operation. Instead, we characterize the quality of the
gate operation by using it for creating different Bell states and
determining their fidelities.

%Bell state analysis

For the Bell state $\Psi_1 = \qdd+i\quu$, the fidelity is given by
$F=\bra{\Psi_1}\rho^{\rm exp}\ket{\Psi_1}=(\rho^{\rm
exp}_{\uparrow \uparrow,\uparrow \uparrow}+\rho^{\rm
exp}_{\downarrow
\downarrow,\downarrow\downarrow})/2+\mbox{Im}\rho^{\rm
exp}_{\downarrow \downarrow,\uparrow \uparrow}$, with the density
matrix $\rho^{\rm exp}$ describing the experimentally produced
state. To determine $F$, we need to measure the populations $p_2 +
p_0$ at the end of the gate operation as well as the off-diagonal
matrix-element $\rho^{\rm exp}_{\downarrow
\downarrow,\uparrow\uparrow}$. To determine the latter, we apply a
$\pi/2$ pulse with optical phase $\phi$ to both ions and measure
$\langle\sigma_z^{(1)}\sigma_z^{(2)}\rangle$ for the resulting
state as a function of $\phi$. This procedure is equivalent to
measuring oscillations of the expectation value
$\mbox{Tr}(P(\phi)\rho^{\rm exp})$ of the operator
$P(\phi)=\sigma_\phi^{(1)}\sigma_\phi^{(2)}$ where
$\sigma_\phi=\sigma_x\cos\phi+\sigma_y\sin\phi$ (see
\fref{fig:hotgate}~(b) and (d)). The amplitude $A$ of these
oscillations equals $2 |\rho^{\rm exp}_{\downarrow
\downarrow,\uparrow\uparrow}|$ and is obtained by fitting them
with the function $P_{\rm fit}(\phi)=A\sin(2\phi+\phi_0)$.

%% Faster gates
Previous measurements \cite{Benhelm:2008b} using \qdd \ as input
state have demonstrated Bell state fidelities as high as 0.993(1)
(see \fref{fig:hotgate} (a) and (b)) for gate times of
$\unit[50]{\mu s}$ or $61$ trap oscillation periods. By doubling
the detuning to $\epsilon/(2\pi)=\unit[40]{kHz}$ we reduce the
gate duration to only $31$ trap oscillation periods and observe
Bell states with a fidelity of $0.971(2)$ which is remarkable
considering the small Lamb-Dicke parameter of $\eta=0.044$. The
detrimental effects illustrated in \fref{fig:offresexc} (a) are
sufficiently suppressed by amplitude pulse shaping.

%% Other input states

Moreover, for a gate time of $\unit[50]{\mu s}$ the analysis was
extended by applying the gate to the state \qdu \ which is
prepared by a $\pi/2$ rotation (beam 2) of both ions, followed by
a $\pi$ phase shift pulse on a single ion performed with the
far-detuned focused beam~1, and another $\pi/2$ rotation applied
to both ions. This pulse sequence realizes the mapping
\begin{equation}\label{eqn:preseq}
\qdd\longrightarrow
|\!\downarrow+\uparrow\rangle|\!\downarrow+\uparrow\rangle\longrightarrow
|\!\downarrow-\uparrow\rangle|\!\downarrow+\uparrow\rangle\longrightarrow\qdu
\end{equation}
to the desired input state for the gate. Imperfections of
single-ion addressing lead to an error in state preparation of
$0.036(3)$. For the Bell state analysis, we measure the population
$p_1$ to infer $\rho^{\rm exp}_{\uparrow \downarrow,\uparrow
\downarrow}+\rho^{\rm exp}_{\downarrow \uparrow,\downarrow
\uparrow}$. Unfortunately parity oscillations cannot be introduced
by a collective $pi/2$ pulse acting on the state  $\qud + i \qdu$.
Instead, we transform this state into $\quu + i\qdd$ by repeating
the steps of sequence (\ref{eqn:preseq}) as for the state
preparation and measure again the coherence by performing parity
oscillations.

\begin{figure}
\centering
\includegraphics[width=80mm]{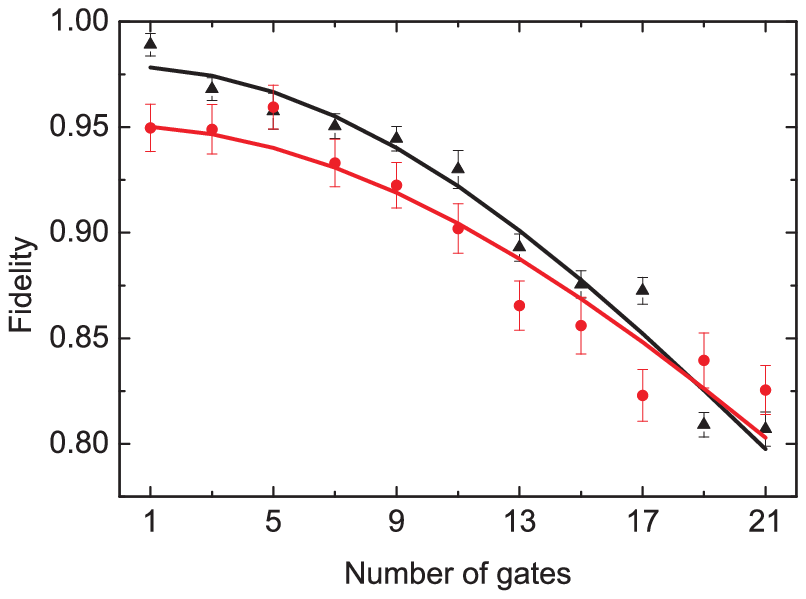}
\caption{\label{fig:multiple_gates} Bell state fidelities after
gate operations applied to the input states \qdd($\fulltriangle$)
and \qdu($\textcolor[rgb]{1,0.00,0.00}{\bullet}$) for $t_{\rm
gate}=50\mu$s. Taking into account the error for state preparation
of the input states $\qdu$ and a similar error to measure the
parity signal, we conclude that the gate operation works on all
tested input states similarly well. The solid lines reflect a
Gaussian decay of the parity fringe amplitudes as a function of
the number of gates and a linear decay in the desired populations
caused by the spectral impurity of the laser. For both input
states the gate operation implies errors of less than $0.2$ after
$21$ consecutive applications.}
\end{figure}

\Fref{fig:multiple_gates} shows a comparison of the fidelity of
the gate starting either in $\qdd$ or $\qdu$. The fidelity of a
Bell state created by a single gate starting in $\qdu$ is
$0.95(1)$. Taking into account the errors for state preparation
and the Bell state analysis we conclude that the entangling
operation works equally well for $\qdu$ as an input state. This
hypothesis is supported by the observation that for both states we
obtain a similar decay of Bell state fidelities with increasing
gate number.

Compared with our earlier results~\cite{Benhelm:2008b} where
multiple gate operations were induced by varying the duration of a
single bichromatic pulse, here we applied up to $21$ individual
amplitude-shaped pulses. Splitting up a long pulse into many
shorter gate pulses has no detectable effect on the fidelity of
the Bell states produced, and in both cases we obtain a Bell state
fidelity larger than $0.80$ after $21$ gates.

\subsection{Gate errors}

\begin{figure}
\centering
\includegraphics[width=80mm]{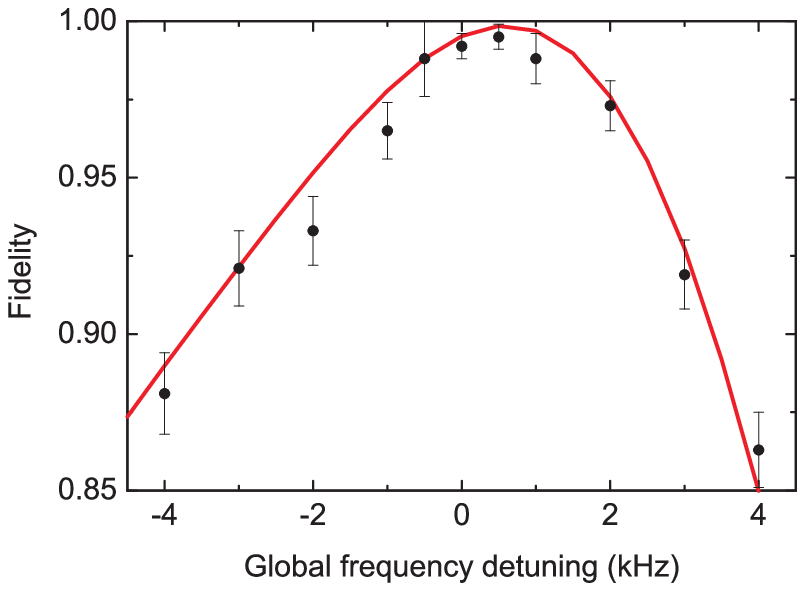}
\caption{\label{fig:global_detuning} Fidelity as a function of the
global frequency detuning of the bichromatic light pulse for from
the carrier transition (here, the sideband detuning was set to
$\epsilon/(2\pi)=\unit[20]{kHz}$). A maximum fidelity of
$0.995(4)$ was found for a detuning of \unit[500]{Hz} from the
transition center due to a residual AC-Stark shift. The solid line
is obtained by numerically solving equation
(\ref{FullHamiltonian}) and taking into account the AC-Stark shift
compensation by different powers of the blue and the red laser
frequency component. At the maximum, the solid line experiences a
second order frequency dependence of $\unit[-9.6(3) \times
10^{-9}]{Hz^2}$.}
\end{figure}

As discussed in~\cite{Benhelm:2008b}, the two dominant sources of
gate errors are laser frequency noise and variations of the
laser-ion coupling strength. Imperfections of the laser's
frequency spectrum lead to incoherent carrier excitation and thus
to a loss in the Bell state fidelity of $2\times10^{-3}$ per gate.
Coupling strength variations of
$\delta\Omega/\Omega\approx1.4(1)\times10^{-2}$ are the major
cause for the Gaussian decay of the parity oscillation amplitudes.

An error that was not investigated in detail before is the
dependence of the Bell state fidelity on the global laser
frequency detuning from the qubit transition frequency.
Experimental results are shown in \fref{fig:global_detuning}. The
solid line fitting the data is calculated by numerically solving
the full Schr{\"o}dinger equation for different global frequency
detunings and evaluating the fidelity. A second order frequency
dependence of $\unit[-9.6(3) \times 10^{-9}]{Hz^2}$ is found from
calculations at the maximum point. This suggests that our laser's
typical mean frequency deviation of $\unit[160]{Hz}$ contributes
with $3\times 10^{-4}$ to the error budget.

A further error source arises when the bichromatic beam couples to
both ions with different strengths. By recording Rabi oscillations
simultaneously on the two ions we conclude that both ions
experience the same coupling strength $\Omega$ to within $4\%$.
From numerical calculations we infer an additional error in the measured
Bell state fidelity of less then $1 \times 10^{-4}$.

Another possible error source is heating of the COM-mode during
the gate operation since the gate is not insensitive to motional
heating in the parameter regime of our implementation.
Using the calculation performed in \cite{Soerensen2000}, we
find a fidelity reduction of $\Delta F=\Gamma_h t_{\rm gate}/2$
where $\Gamma_h$ is the heating rate of the COM-mode. As in our
experiments $\Gamma_h = 3s^{-1}$, the fidelity is reduced by
$\Delta F\approx 10^{-4}$ for $t_{\rm gate}=\unit[50]{\mu s}$.

\section{A M{\o}lmer-S{\o}rensen gate with ions in a thermal state of motion}

\subsection{Formal description of the time evolution}

In theory, the M{\o}lmer-S{\o}rensen gate does not require the
ions to be cooled to the ground states of
motion since its propagator (\ref{MSPropagator}) is independent of
the vibrational state for $t=t_{\rm gate}$. For $t\neq t_{\rm
gate}$, however, the interaction entangles qubit states and
vibrational states so that the qubits' final state becomes
dependent on the initial vibrational state. Therefore, it is of
interest to calculate expectation values of observables acting on
the qubit state space after applying the propagator for an
arbitrary time $t$. As will be shown below, simple expressions can
be derived in the case of a thermally occupied motional state. For
the following calculation, it is convenient to define
$V(t)=\exp(i\gamma S_y^2)$ where $\gamma = \lambda t
-\chi\sin(\epsilon t)$. We are interested in calculating the
expectation value of the observable $\cal{O}$ given by
\begin{eqnarray}
O(t)&=& \mbox{Tr}({\cal{O}}U(t)\rho_M\otimes\rho_A U(t)^\dagger)\nonumber\\
&=&
\mbox{Tr}(\rho_M\otimes\rho_A\hat{D}(-\alpha S_y)V^\dagger {\cal{O}}V\hat{D}(\alpha S_y))\nonumber\\
&=&
\sum_n p_n \mbox{Tr}_A(\rho_A\langle n|\hat{D}(-\alpha
S_y)V^\dagger{\cal{O}}V\hat{D}(\alpha S_y)|n\rangle). \label{Ooft}
\end{eqnarray}
Here, $\rho_M=\sum_n p_n |n\rangle\langle n|$ with
$p_n=\frac{1}{\bar{n}+1}\left(\frac{\bar{n}}{\bar{n}+1}\right)^n$
describes a thermal state with average phonon number $\bar{n}$ and
$\mbox{Tr}_A$ denotes the trace over the qubit state space with
$\rho_A$ being the initial state of the qubits. For two ions, the
state-dependent displacement operator $\hat{D}(\alpha S_y)$ is
given by
\[
\hat{D}(\alpha S_y) = P_0 + P_2 \hat{D}(2\alpha) + P_{-2}
\hat{D}(-2\alpha),
\]
where $P_\lambda$ is the projector onto the space spanned by the
eigenvectors of $S_y$ having the eigenvalue $\lambda$, with
$P_0=\mathbbm{1}-\frac{1}{4}S_y^2$, and $P_{\pm
2}=\frac{1}{8}(S_y^2\pm 2S_y)$. This decomposition allows for
tracing over the vibrational states in (\ref{Ooft}) since $\langle
n|\hat{D}(\alpha)|n\rangle =
\exp(-|\alpha|^2/2){\cal{L}}_n(|\alpha|^2)$, where ${\cal{L}}_n$
denotes a Laguerre polynomial. For taking the trace, we note that
$\sum_n p_n \langle n|\hat{D}(\alpha)|n\rangle$ is proportional to
the generating function $g(x,\beta)$ of the Laguerre polynomial
\cite{Abramowitz:1972_p784} given by
\[
g(x,\beta)=\sum_{n=0}^\infty {\cal{L}}_n(\beta)
x^n=\frac{1}{1-x}\exp\left(-\frac{\beta x}{1-x}\right).
\]
%(see Abramowitz and Stegun, chapter 22.9)
Therefore,
\begin{equation}
\label{nDnThermal} \sum_n p_n \langle n|\hat{D}(\alpha)|n\rangle=
\frac{1}{\bar{n}+1}\,g(\frac{\bar{n}}{\bar{n}+1},|\alpha|^2)\,\exp(-|\alpha|^2/2)
=e^{-|\alpha|^2(\bar{n}+\frac{1}{2})}.
\end{equation}
 Using the abbreviation ${\cal{O}_V}=V{\cal{O}}V^\dagger$ and
(\ref{nDnThermal}), the expectation value $O(t)$ is given by
\begin{equation}
O(t)=\mbox{Tr}_A({\cal{O}_V}\{ A_0 + A_4
e^{-4|\alpha|^2(\bar{n}+\frac{1}{2})}+A_{16}
e^{-16|\alpha|^2(\bar{n}+\frac{1}{2})}\})
\end{equation}
with
\begin{eqnarray}
A_0&=&P_0\rho_A P_0+P_2\rho_A P_2+P_{-2}\rho_A P_{-2}\nonumber\\
A_4&=&P_2\rho_A P_0+P_0\rho_A P_2+P_{-2}\rho_A P_0 + P_0\rho_A P_{-2}\nonumber\\
A_{16}&=&P_{-2}\rho_A P_2 + P_2\rho_A P_{-2}\,.\nonumber
\end{eqnarray}
For the initial state
$\rho_A=|\downarrow\downarrow\rangle\langle\downarrow\downarrow\!|$,
one obtains
\[
A_0=(S_z^2+S_x^2)/16,\;\;A_4=-S_z/4,\;\;A_{16}=(S_z^2-S_x^2)/16
\]
and
\[
O(t)=\frac{1}{16}\mbox{Tr}_A({\cal{O}_V}\{(S_z^2+S_x^2) -4S_z
e^{-4|\alpha|^2(\bar{n}+\frac{1}{2})}+(S_z^2-S_x^2)
e^{-16|\alpha|^2(\bar{n}+\frac{1}{2})}\}).
\]
To calculate the time evolution of the qubit state populations
starting from state $|\!\downarrow\downarrow\rangle$ at $t=0$, use
of the relations
\begin{eqnarray*}
e^{-i\gamma S_y^2}S_ze^{i\gamma S_y^2}&=&\cos(4\gamma)S_z-\sin(4\gamma)\frac{1}{2}\{S_x,S_y\}\\
e^{-i\gamma S_y^2}S_j^2e^{i\gamma S_y^2}&=&S_j^2
%\;\;\; \mbox{for} \;\gamma\in\mathbb{R}.
\end{eqnarray*}
yields the following expressions for the qubit state populations:
\begin{eqnarray}
p_2(t)&=&
\frac{1}{8}(3+e^{-16|\alpha|^2(\bar{n}+\frac{1}{2})}+4\cos(4\gamma)e^{-4|\alpha|^2(\bar{n}+\frac{1}{2})})\nonumber\\
p_1(t)&=&
\frac{1}{4}(1-e^{-16|\alpha|^2(\bar{n}+\frac{1}{2})})\label{thermalpops}\\
p_0(t)&=&
\frac{1}{8}(3+e^{-16|\alpha|^2(\bar{n}+\frac{1}{2})}-4\cos(4\gamma)e^{-4|\alpha|^2(\bar{n}+\frac{1}{2})})\nonumber
\end{eqnarray}
The formalism presented here could also be used to calculate the
contrast of a parity scan for thermal states of motion. In this
case, the parity operator is given by $P=S_z^2/2-\mathbbm{1}$. The
$\pi/2$ carrier pulses transform this operator into an operator
$P_\phi=(\cos\phi S_x+\sin\phi S_y)^2/2-\mathbbm{1}$.

\subsection{High fidelity Bell states of ions in a thermal state}

While many theoretical papers discussing M{\o}lmer-S{\o}rensen and
conditional phase gates put much emphasis on the possibility of
entangling ions irrespective of their motional state by using
these gates, there has not been any experimental demonstration of
this gate property up to now. The reason for this is that
independence of the motional state, as predicted by
(\ref{MSPropagator}), is achieved only deep within the Lamb-Dicke
regime whereas experiments demonstrating entangling gates on
hyperfine qubits usually have Lamb-Dicke factors on the order of
$\eta=~0.1-0.2$ \cite{Sackett2000,Leibfried2003a,Haljan:2005b}.
% (Lamb-Dicke parameters: Sackett00: 0.16 for two ions, Leibfried03: 0.19, Haljan05: 0.1)
Therefore, all previous experimental gate realizations used laser
cooling to prepare at least the motional mode mediating the gate
in its ground state with $n=0$.

\begin{figure}
\centering
\includegraphics[width=140mm]{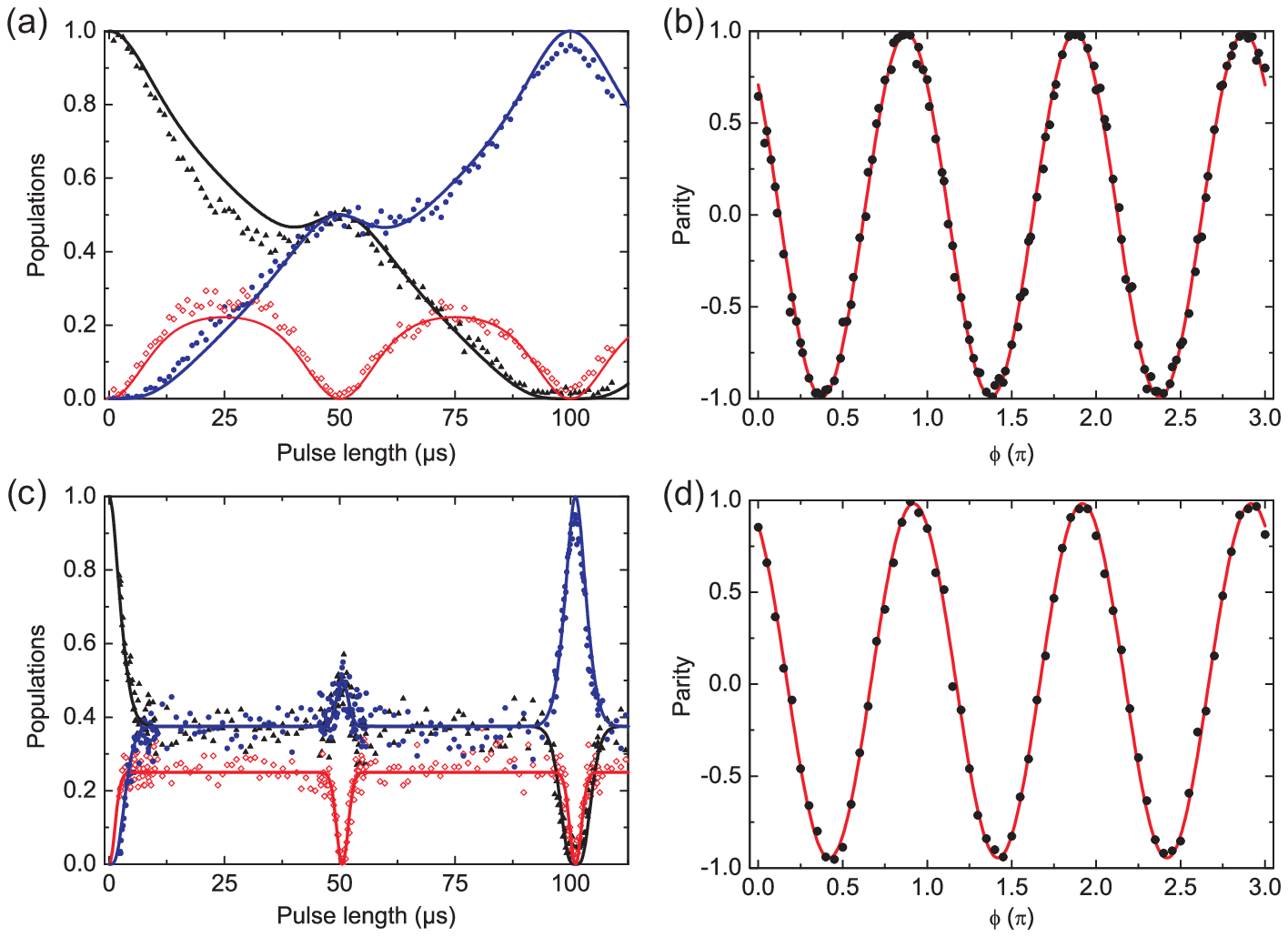}
\caption{\label{fig:hotgate} Measured population evolution for
$p_0 (\blacktriangle), p_1
(\textcolor[rgb]{1.00,0.00,0.00}{\diamond}),
p_2(\textcolor[rgb]{0.00,0.00,1.00}{\bullet}) $ and parity
oscillations with (a,b) and without (c,d) ground state cooling. In
the latter case, population is transferred faster into
\ket{\mbox{$\uparrow \downarrow, n$}}, \ket{\mbox{$\downarrow
\uparrow, n$}} as compared to sideband cooled ions due to the
higher coupling strength to the sidebands. In (c), the solid lines
are a fit to the data points using~(\ref{thermalpops}) with the
mean phonon number $\bar{n}$ as a free parameter giving
$\bar{n}=20(2)$. The parity oscillations for the ions in a thermal
state of motion have an amplitude of $0.964(2)$. Combining both
measurements, we determine the Bell state fidelity to be
$0.974(1)$. The data appearing in (a) and (b) are taken
from~\cite{Benhelm:2008b}. Here the deviation of the solid lines
from the data is caused by the AC-Stark shift compensation using
$\zeta=0.08$. }
\end{figure}

\Fref{fig:hotgate}~(a) illustrates the population evolution induced by
the gate pulse for ground-state cooled ions initially prepared in
the qubit states \qdd, \fref{fig:hotgate}~(b) displays parity
oscillations for the produced Bell state. The corresponding time evolution and
parity oscillations for ions that are merely Doppler-cooled to a
thermal state with $\bar{n}=20(2)$ are shown in
\fref{fig:hotgate}~(c) and (d) respectively. As the coupling
strengths on the upper and lower motional sidebands scale
 as $\propto\sqrt{n+1}$ and $\propto\sqrt{n}$, non-resonant sideband excitation transfers
population much faster from $\ket{\!\!\downarrow \downarrow,n}$
into $\ket{\!\!\downarrow \uparrow,n\pm 1}$, $\ket{\!\!\uparrow
\downarrow,n\pm 1}$ as compared to the case of ions prepared in
the ground state with $\bar{n}=0$. After the gate time $t_{\rm
gate}=\unit[50]{\mu s}$, however, the undesired population $p_1$
nearly vanishes as in the case of ground-state cooled ions and the
Bell state $\Psi_1$ is again created. In the experiment, we find a
population $p_1=0.015(1)$ in the undesired energy eigenstates. The
parity oscillations have an amplitude of $0.964(2)$, resulting in
a Bell state fidelity of $0.974(1)$. The reasons for the somewhat
reduced fidelity as compared with ground-state cooled ions are
currently not well understood. In part, the fidelity loss arises from a variation
of the coupling strength on the vibrational sidebands as a
function of $n$ caused by higher-order terms in $\eta$. However,
for a thermal state with $\bar{n}=20$ and $\eta=0.044$
calculations show this effect amounts only to additional errors of
$7\times 10^{-3}$.

As mentioned in Section~\ref{AC_stark}, the AC-Stark compensation
by unbalancing the power of the red and blue frequency component
is not applicable to ions in a thermal state. Instead, the laser
frequency needs to be adjusted to account for AC-Stark shifts
$\delta_{\rm AC}$, a technique that works well as long as the
AC-Stark shifts are smaller than the coupling strength $\lambda$
of the gate interaction appearing in (\ref{Heff}) (otherwise, in
the case $\delta_{\rm AC}\gg\lambda$, small laser power
fluctuations give rise to large phase shifts). Therefore, care
must be taken to choose the direction and polarization of the gate
laser such that a favorable ratio $\lambda/\delta_{\rm AC}$ is
obtained. In experiments with a gate duration of $t_{\rm gate}=\unit[50]{\mu s}$, we achieved $\lambda/\delta_{\rm
AC}\approx 3$ and needed to shift the laser frequency by about
\unit[7.5]{kHz} for optimal Bell state fidelity. In future experiments, a further
reduction of the AC-Stark shift could be obtained using a
$\sigma^+$-polarized laser beam incident on the ions along the
direction of the magnetic field. In this geometry the AC-Stark
shift is predominantly caused by the $S_{1/2}\leftrightarrow
P_{3/2}$ dipole transition since the $D_{5/2}(m=+3/2)$ state does
not couple to any of the 4p Zeeman states. From calculations we
infer a reduction of the shift to about \unit[2]{kHz} without
compromising the gate speed.

Fitting equations (\ref{thermalpops}) to the population evolution
data allows us to determine the mean vibrational quantum number as
$\bar{n}=20(2)$. This value is
consistent with independent measurements obtained by comparing the
time evolution of the ions when exciting them on the carrier and
on the blue motional sideband.

\section{Conclusions and outlook}

Until recently, entangling gates for optical qubits were
exclusively of the Cirac-Zoller type which require individual
addressing of the ions. Compared to this type of gate the
M{\o}lmer-S{\o}rensen gate gives an improvement in fidelity and
speed of nearly an order of magnitude. The achieved fidelity sets
a record for creating two-qubit entanglement on demand
irrespective of the physical realization considered so far. Our
results with concatenations of $21$ of these operations bring the
realization of more complex algorithms a step closer to reality.
The implementation of a gate without the need for ground state
cooling is of particular interest in view of quantum algorithms
that require entangling gates conditioned on quantum state
measurements that do not preserve the ions' motional quantum
state.

The optical qubit as used here is certainly not the best solution
for long time storage of quantum information. Instead, qubits
encoded in two hyperfine ground states whose frequency difference
is insensitive to changes in magnetic field are preferable. These
magnetic-field insensitive hyperfine qubits can store quantum
information for times exceeding the duration of the gate operation
presented here by more than four orders of
magnitude~\cite{Langer2005,olmschenk2007,Benhelm:2008c}. However,
on such qubit states no high-fidelity universal gates have been
demonstrated so far. Hence, our next experimental efforts will
focus on implementing the M{\o}lmer-S{\o}rensen gate using
$\mathrm{^{43}Ca^+}$ ions. By mapping between the hyperfine qubit
encoded in the ion's ground states and the optical qubit we will
benefit from both of their advantages.

Another interesting perspective of this gate is to create
multi-qubit interactions between more than two qubits. A gate
collectively interacting with all ions at the same time, in
combination with a collective spin flips and a strongly focused
off-resonant laser beam inducing phase shifts in individual ions,
constitutes a basis set of Hamiltonians that offers the prospect
of realizing complex multi-qubit operations such as a Toffoli-gate
and a quantum error-correcting algorithm~\cite{Nebendahl:2008mip}.

%\begin{acknowledgments}
\ack We gratefully acknowledge the support of the European network
SCALA, the Institut f{\"u}r Quanteninformation GmbH and IARPA.
R.~G. acknowledges funding by the Marie-Curie program of the
European Union (grant number PIEF-GA-2008-220105).
%\end{acknowledgments}

\section*{References}

\bibliographystyle{iopart-num}
%\bibliographystyle{unsrt}

%\bibliography{biblio-most-recent}

\providecommand{\newblock}{}

\end{document}